\documentclass{article}
\usepackage{PRIMEarxiv}
\usepackage[utf8]{inputenc} 
\usepackage[T1]{fontenc}    
\usepackage{hyperref}       
\usepackage{url}            
\usepackage{booktabs}       
\usepackage{amsfonts}       
\usepackage{nicefrac}       
\usepackage{microtype}      
\usepackage{lipsum}
\usepackage{fancyhdr}       
\usepackage{graphicx}       
\graphicspath{{media/}}     

\usepackage{amsmath,amssymb,array}
\usepackage{booktabs}

\title{Good distribution modelling with the \texttt{R} package \texttt{good}}

\author{
  Jordi Tur, David Moriña, Pedro Puig, Alejandra Cabaña, Argimiro Arratia, Amanda Fernández-Fontelo
}

\begin{document}
\maketitle

\begin{abstract}
Although models for count data with over-dispersion have been widely considered in the literature, models for under-dispersion -the opposite phenomenon- have received less attention as it is only relatively common in particular research fields such as biodosimetry and ecology.
The Good distribution is a flexible alternative for modelling count data showing either over-dispersion or under-dispersion, although no \texttt{R} packages are still available to the best of our knowledge. We aim to present in the following the \texttt{R} package \texttt{good} that computes the standard probabilistic functions (i.e., probability density function, cumulative distribution function, and quantile function) and generates random samples from a population following a Good distribution. The package also considers a function for Good regression, including covariates in a similar way to that of the standard \texttt{glm} function. We finally show the use of such a package with some real-world data examples addressing both over-dispersion and especially under-dispersion.
\end{abstract}


\section{Introduction}
Even though the Poisson distribution is the most broadly used for modelling count data (e.g., properties like closure under addition make this distribution especially useful in practice) and has been used to model applications such as the number of soldiers killed by horse kicks per Prussian army corp (\cite{vB98}) or the number of chromosome aberrations (e.g., dicentrics or rings) in scored sample cells from potentially irradiated individuals (\cite{IAEA}), it is usually chosen for tradition and convenience, and such a practice can very often be brought into question. The Poisson distribution assumes a ratio of the mean to the variance (i.e., dispersion index) of one (i.e., equi-dispersed distribution with equal mean and variance), which is usually not satisfied in real-world data settings, leading thus to bad data fittings and biased parameter estimates. The Poisson distribution is neither a good alternative if, e.g., our data setting exhibits zero-inflation, as such a phenomenon is best addressed with two-parameter distributions according to \cite{P03}. 

Deviations from the Poisson assumption (i.e., equi-dispersion) leads to alternative discrete-valued (count) distributions that can either fit over-dispersion (i.e., the dispersion index is over one) or (and) under-dispersion (i.e., the dispersion index is under one). These distributions are mostly focused on fitting over-dispersion, as such a phenomenon is rather common in count data, but some can also model under-dispersion. The literature gives an extensive array of count distributions that fit over-dispersion, such as the Negative Binomial distribution (\cite{GJ20,CPC05, H08}), the generalized Poisson (GP) distribution (\cite{C89}), the double Poisson (DP) distribution (\cite{E86}), and the $\textrm{r}^{\textrm{th}} $ Hermite distribution (\cite{P03, M15}). We can also model over-dispersion through, e.g., a mixture of Poisson distributions and capture zero-inflation through, e.g., a zero-inflated Poisson distribution. See \cite{JKK}, Chapters 8.2.5 and 4.10.3.

We focus here on under-dispersion with respect to the Poisson distribution, which is present in particular research fields such as biodosimetry (\cite{P14}) or community ecology (\cite{H14}) and has been further properly addressed by authors such as \cite{S12} and \cite{BW17}; however, such a phenomenon is relatively less common in count data and only few probability distributions can model such a phenomenon. In particular, under-dispersion has been addressed before with discrete-valued distributions such as the Conway-Maxwell (COM) Poisson distribution (\cite{CM62, C89, S12}), DP distribution (\cite{E86}) or GP distribution (\cite{C89}). However, these distributions show some limitations that could sometimes cause further difficulties. The two-parameter COM-Poisson distribution, for example, models under-dispersion and over-dispersion depending on whether the parameter $\nu$ (dispersion parameter) is $\nu \in (1,+\infty)$ (under-dispersion) or $\nu \in [0,1)$ (over-dispersion) (\cite{K08}). One of the main limitations of the COM-Poisson distribution relies on its probability mass function's normalizing constant (and therefore related characteristics such as the expectation, variance and higher moments), as it is approximated numerically, and thus requires expensive computations very often. For further details see \cite{S05}, \cite{W13a}, \cite{LI19}, \cite{S12}, and some of the references therein. On the other hand, the two-parameter DP distribution also allows for under-dispersion ($\psi>1$) and over-dispersion ($\psi<1$). See \cite{JKK} for further details on this distribution and its parametrization. As for the COM-Poisson distribution, the DP probability mass function's normalizing constant is computed numerically, meaning that its characteristics (e.g., expectation) are only known approximately (\cite{W13a}). Finally, for the two-parameter GP distribution (also called Lagrangian Poisson distribution), both under-dispersion and over-dispersion are fitted depending on whether the parameter $\lambda<0$ or $\lambda >0$, respectively. Although the properties of the distribution are completely known when $0<\lambda<1$ (over-dispersion), \cite{N75} pointed out that negative probabilities could be obtained when $\lambda <0$ (under-dispersion). \cite{CS85} overcame such a problem by proposing a truncation of the GP distribution. This solution, however, implies the distribution support depends on the value at which the distribution is truncated, the probabilities do not sum $1$ exactly anymore (as they are approximated), and the properties of the distribution are satisfied only in an approximate manner (\cite{JKK,C89}). \cite{W13a} additionally stressed that this truncation might cause problems for parameter estimation such as maximum likelihood methods. 

In the following, an alternative count distribution called Good distribution \cite{G53} is considered, especially to address under-dispersion compared to the Poisson distribution. This is a two-parameter distribution, and a particular case of the three-parameter Lerch distribution (\cite{ZA95}), which addresses under-dispersion and over-dispersion compared to a Poisson distribution. Since the Good distribution was proposed by (\cite{G53}) in a paper entitled "The population frequencies of species and the estimation of population parameters", it has been used in ecology applications as discussed by \cite{K92}, and sequential pattern analysis (\cite{WP13}). Interesting applications can also be found in \cite{IS77} and \cite{K95}. As under-dispersion is rather present in applications of some research areas (e.g., biodosimetry or ecology), a broader array of distributions also addressing under-dispersion is thus needed, especially if some existing alternatives might additionally show problems in practice (e.g., problems with maximum likelihood estimation due to the truncation of the GP distribution in case of under-dispersion). Therefore, we present and describe in the current paper a tool in the form of an \texttt{R} package to robustly and efficiently compute useful characteristics of the Good distribution (e.g., probabilities and quantiles) and fit a Good regression which also allows incorporating covariates related to the distribution parameters. 

To formally introduce the Good distribution, we need to define first the three-parameter Lerch distribution (\cite{ZA95}). The probability mass function (pmf) of the Lerch distribution is given by:
\begin{equation}\label{eq1}
\textrm{P}\left(X=x\right) = \frac{z^x\left(\nu+x\right)^{-s}}{\Phi \left(z,s,\nu\right)}, \quad \textrm{for} \quad 0 < z < 1, \quad \nu >0,\quad s \in \mathbb{R}   
\end{equation}
whose support is on the set of the non-negative integers $x = 0, 1, \dots$, and $\Phi \left(z,s,\nu\right)$ is the so-called normalizing constant, which is given by the so-called Lerch’s  transcendent function: 
\begin{equation}\label{eq4}
\Phi \left(q,s,\nu\right) = \sum ^\infty _{n=0} \frac{q^n}{\left(\nu+n\right)^{s}}, \quad \textrm{for} \quad |q| < 1, \quad \nu>0,
\end{equation}

satisfying the relation: $\Phi\left(q,s,\nu\right)=q^m\Phi\left(q,s,m+\nu\right)+\sum_{n=0}^m \frac{q^n}{\left(\nu+n\right)^{s}}$ for $m=1,2,3,\cdots$ (\cite{WOS66}). Notice that we ensure the equation (\ref{eq1}) is always positive by restricting to $z>0$ and $\nu>0$. Distributions such as the discrete Pareto distribution ($z=1$, $s>1$ and $\nu=1$), the Logarithmic distribution ($0<z<1$, $s=1$ and $\nu=1$), the Geometric distribution ($0<z<1$, $s=0$ and $\nu=1$) and Good distribution ($0<z<1$, $s\in \mathbb{R}$, and $\nu=1$) are particular cases of the Lerch distribution. See \cite{JKK} (Chapter 11.2.20) for further particular cases of the Lerch distribution. Notice, however, that \cite{ZA95} used a version of the Lerch distribution with a truncation at $0$.  

The ratio of the variance to the mean of the Lerch distribution can be lower, equal or greater than $1$, and thus this distribution can either fit under-dispersion, equi-dispersion and over-dispersion. Expressions for the expectation, variance and other characteristics of the Lerch distribution can be found in \cite{W13a}, \cite{ZA95} and \cite{K10}. 
A particular case of the Lerch distribution is the so-called Good distribution that results when $s\in \mathbb{R}$, $0<z<1$ and $\nu=1$. Following \cite{W13a}, it can be seen that the Lerch's transcendent function in expression (\ref{eq4}) derives to the so-called polylogarithm function (also called Jonquière's function) when $\nu=1$:
\begin{equation}\label{eq5}
\textrm{F}\left(z,s\right) = \sum^\infty _{n=1} \frac{z^{n}}{n^{s}}=z \cdot \Phi\left(z,s,1\right) \quad \text{for} \ |z| < 1,
\end{equation}
where $\Phi\left(\cdot,\cdot,\cdot\right)$ is the Lerch transcendent function defined in expression (\ref{eq4}). Therefore, the Good distribution is represented by the following pmf that depends on parameters $z$ and $s$:
\begin{equation}\label{eq6}
\textrm{P}\left(X=x\right) = \frac{z^{x+1}\left (x+1\right)^{-s}}{\textrm{F}(z,s)}, \quad \text{for} \ 0<z<1 \text{ and } s \in \mathbb{R}, 
\end{equation}

with support on the set of the non-negative integers $x=0, 1, 2, \cdots$. Notice that \cite{JKK} (with parameters $q=z$ and $\eta=s$), \cite{K92} (with parameters $q=z$ and $\alpha=-s$), and \cite{DL97} (with parameters $q=z$ and $a=-s$), for example, used alternative parametrizations of the Good distribution. \cite{K92} also provided a nice interpretation for the pmf in expression (\ref{eq6}) based on the behaviour of functions $z^{x+1}$ and $(x+1)^{-s}$ and the values of parameters $z$ and $s$, and showed graphically that parameters $s$ and $z$ somehow act as shape and scale parameters respectively. 

\begin{figure}%
    \centering
    \includegraphics[width=6.9cm]{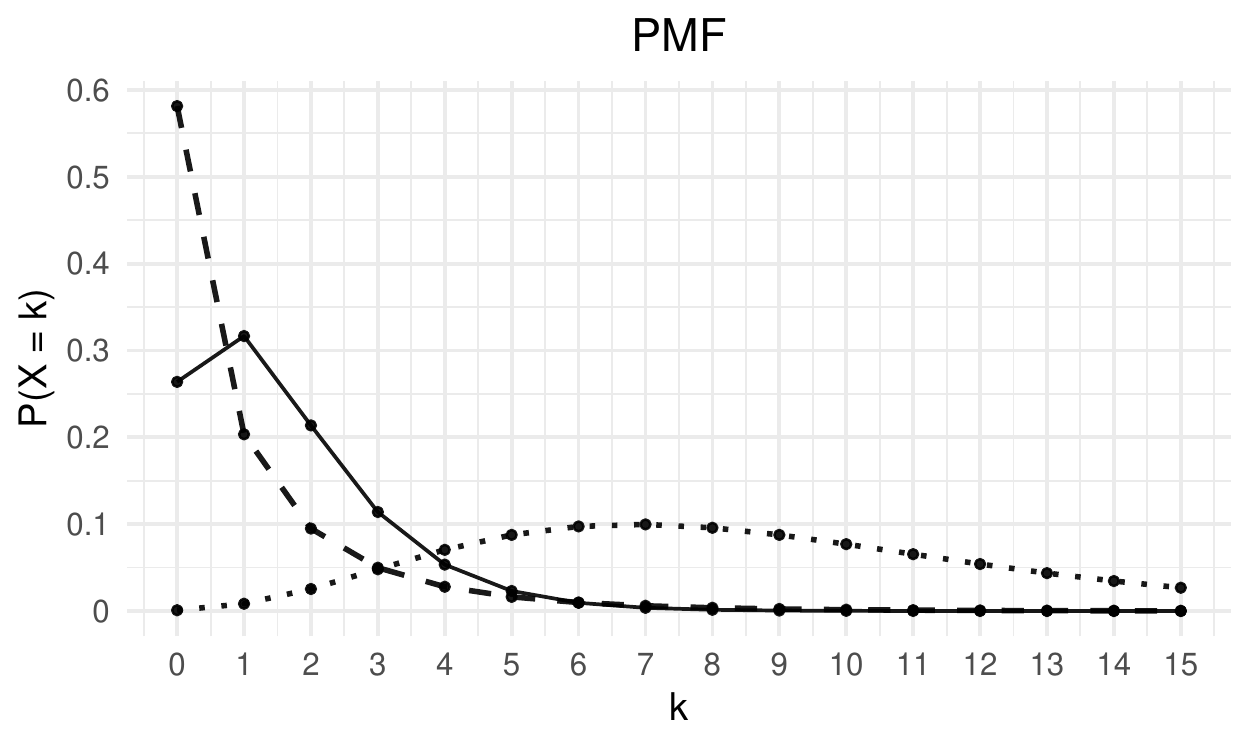} \includegraphics[width=6.9cm]{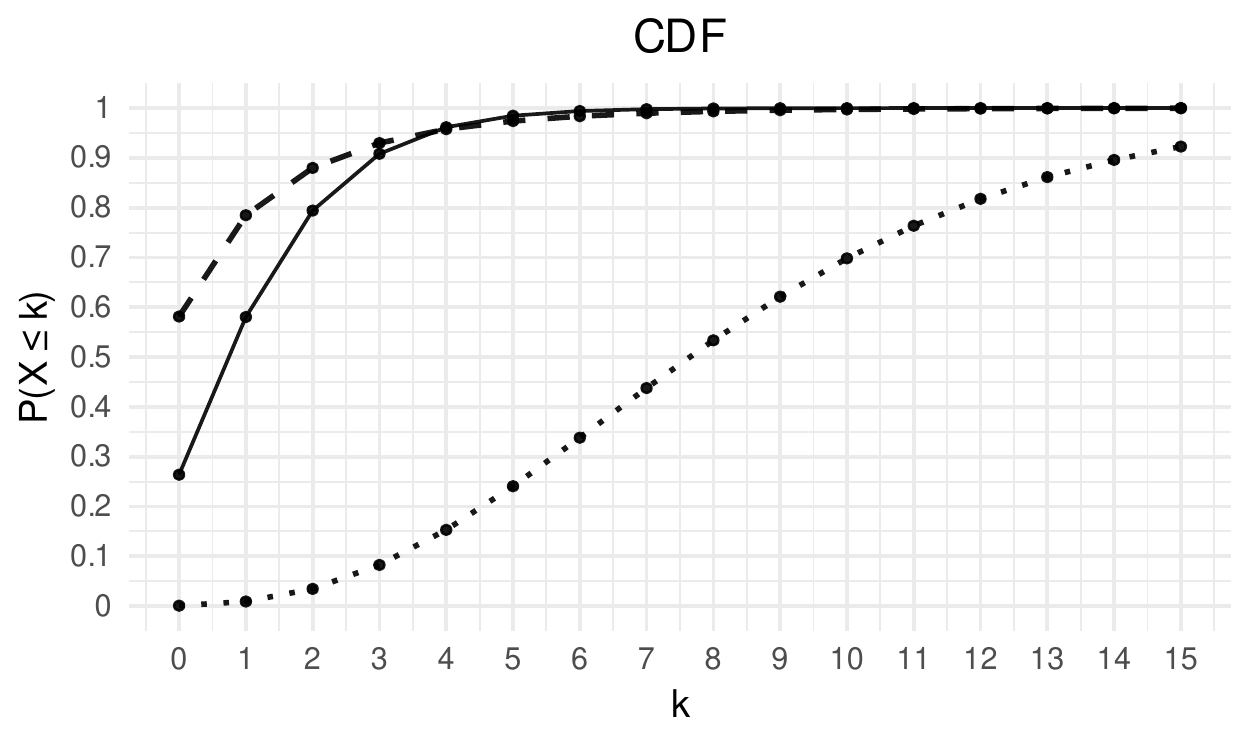}
    \vspace{-0mm}
    \caption{Probability mass function (left) and cumulative distribution function (right) of the Good distribution with parameters $z=0.3, \ s = -2$ (solid), $z=0.7, \ s = 1$ (striped) and $z=0.6, \ s = -4$ (dashed).}%
    \label{fig1}%
\end{figure}

The probability generating function (pgf) of the Good distribution as in expression (\ref{eq6}) is given by:
\begin{equation}\label{eq7}
\textrm{G}_X\left(t\right) = \textrm{E}\left(t^X\right)=\sum_{x=0}^{\infty} t^x\frac{z^{x+1}\left(x+1\right)^{-s}}{\textrm{F}\left(z,s\right)}=\frac{1}{t \textrm{F}\left(z,s\right)}\sum_{x=0}^\infty \frac{\left(zt\right)^{x+1}}{\left(x+1\right)^s}=\frac{1}{t}\frac{\textrm{F}\left(zt,s\right)}{\textrm{F}\left(z,s\right)},
\end{equation}

and its moment generating function (mgf) is given by:
\begin{align}\label{mgf}
\textrm{M}_X\left(t\right)&=\textrm{E}\left(\exp\left(tX\right)\right)=\sum_{x=0}^{\infty} \exp\left(tx\right)\frac{z^{x+1}\left(x+1\right)^{-s}}{\textrm{F}\left(z,s\right)}=\frac{1}{\exp\left(t\right) \textrm{F}\left(z,s\right)}\sum_{x=0}^{\infty} \frac{\left(\exp\left(t\right) z\right)^{x+1}}{\left(x+1\right)^s} \nonumber \\&=\frac{1}{\exp\left(t\right)}\frac{\textrm{F}\left(z\exp\left(t\right), s\right)}{\textrm{F}\left(z,s\right)}.
\end{align}

From both the pgf and mgf in equations (\ref{eq7}) and (\ref{mgf}), the moments of the Good distribution can be derived easily. Expressions for the expectation and variance are given below:
\begin{equation}\label{eq9}
\textrm{E}\left(X\right) = \mu= \frac{\textrm{F}\left(z,s-1\right)}{\textrm{F}\left(z,s\right)} - 1, \quad
\textrm{V}\left(X\right) = \sigma^2= \frac{\textrm{F}\left(z,s-2\right)}{\textrm{F}\left(z,s\right)} -\frac{\textrm{F}^2\left(z,s-1\right)}{\textrm{F}^2\left(z,s\right)},
\end{equation}

and also a general expression for $\textrm{E}(X^k)$ can be derived from the mgf. In particular:
\begin{equation}\label{EXk}
\textrm{E}\left(X^k\right)=\textrm{M}_X^{(k)}(0)=\frac{1}{\textrm{F}(z,s)}\sum_{m=0}^k (-1)^{m+k}\binom{k}{m} \textrm{F}\left(z,s-m\right),
\end{equation}

The expression (\ref{EXk}) is also derived by \cite{Kemp1999}. Notice that as \cite{K92} defined the Good distribution over $x=1, 2, \cdots$ compared to our definition here over $x=0, 1, 2, \cdots$, the expressions for expectation and variance slightly vary. Notice also that the function $\textrm{b}(q,\alpha)$ in \cite{K92} is equivalent to $\textrm{F}(z,-s)$ given in equation (\ref{eq5}). 

The ratio of the variance to the mean thus takes the form:
\begin{equation}
\frac{\sigma^2}{\mu}=\frac{\textrm{F}\left(z,s\right)\textrm{F}\left(z,s-2\right)-\textrm{F}^2\left(z,s-1\right)}{\textrm{F}\left(z,s\right)\textrm{F}\left(z,s-1\right)-\textrm{F}^2\left(z,s\right)},  
\end{equation} 

which can be lower, equal or greater than $1$. We have empirically studied the behaviour of the dispersion index depending on the values of parameters $z$ (e.g., from $0$ to $1$) and $s$ (e.g., from $-10$ to $10$), mainly showing that under-dispersion is related to small values of parameter $z$ (e.g., below $0.20$) with negative values of $s$ (e.g., roughly between $-10$ and $-5$). However, over-dispersion seems to be mainly present for values of the parameter $z$, e.g., over $0.20$ and negative and positive values for the parameter $s$, e.g., between $-10$ and $6$.  

\cite{DL97} proposed an alternative parametrization of the Good distribution that has been previously used by other authors, e.g., \cite{W13a}. The authors provided this new parametrization in the context of parameter estimation; in particular, they used such a distribution to introduce a quadratic distance estimator for parameters $z$ and $s$ and study the asymptotic estimator properties. The parametrization essentially uses the idea of defining a power function through natural logarithms such that:
\begin{equation}\label{eq9.5}
\textrm{P}\left(X=x\right) = \frac{\exp\left(\log(z)(x+1) -s\log(x+1)\right)}{\text{F}\left(\log(z),s\right)}, \quad \text{with} \ \log(z) < 0 \text{ and } s \in \mathbb{R}, 
\end{equation}

where $\textrm{F}\left(\log(z),s\right)=\sum_{n=1}^{\infty} \exp\left(n\log(z)-s\log(n)\right)$. In the reminder of this article, we use the parametrization above for parameter estimation. Notice that in \cite{DL97} parameter $\alpha$ equates $\log(z)$ and parameter $\beta$ equates $-s$. The authors also pointed out that the pdf of the Good distribution as given in equation (\ref{eq9.5}) remains a discrete version of a Gamma distribution with parameters $s$ and $z$ being respectively the shape and scale parameters, although for a Gamma distribution both parameters should be positive, and this is not necessarily the case for the Good distribution. Notice also that the latter is also consistent with the interpretation of the parameters given by \cite{K92}. The \texttt{good} package has implemented this parametrization for likelihood-based parameter estimation. 

This paper is organized as follows. A description of the use of the \texttt{good} package is given in Section \nameref{pack}. Several examples of application are discussed in Section \nameref{exp}, and some conclusions  are derived in Section \nameref{con}.

\section{Package \texttt{good}} \label{pack}

The same as for other probability distributions already implemented in \texttt{R}, the package \texttt{good} includes functions for the pmf  (\texttt{dgood}), the cumulative distribution function (cdf) (\texttt{pgood}), the quantile function (\texttt{qgood}) and a function for random number generation (\texttt{rgood}) for the Good distribution. The package also contains the function \texttt{glm.good}, which works the same way as the standard function \texttt{glm}, to estimate a linear regression by maximum likelihood with a response following a Good distribution and the covariates being either numerical or factors.

\subsection{Probability mass function}

The pmf of the Good distribution is implemented in \texttt{good} through the function \texttt{dgood}. This function can be called as given below: 

\begin{example}
dgood ( x , z , s )
\end{example}

The description of these arguments is summarized as follows:
\begin{itemize}
\item \texttt{x}: vector of non-negative integer values.
\item \texttt{z}: vector of first parameter of the Good distribution. 
\item \texttt{s}: vector of second parameter of the Good distribution.
\end{itemize}

The expression (\ref{eq5}) is computed with the function \texttt{polylog} from the package \texttt{copula} (\cite{copula}). Notice that this function allows computing the polylogarithm function for numeric and complex values for parameter $z$ and integer and real values for parameter $s$. However, the function \texttt{polylog} mainly fails for extreme cases with large negative values of the parameter $s$, e.g., roughly below $-100$, as in these cases the terms $z^{n+1}$ and $(n+1)^s$ multiply, but the term $(n+1)^s$ approaches to infinity dramatically fast. This problem implies that no probabilities can be given for extreme cases, which is, of course, not ideal. We have thus considered an approximation of the polylogarithm function for these extreme cases, and, in particular, we have used the approximation given by \cite{Wood}. The author re-writes the polylogarithm function given in expression (\ref{eq5}) as a sum of two components defined as (see expressions 9.3 and 9.6 in \cite{Wood}):
\begin{align}\label{eq:approx}
\textrm{F}(\exp(z),s)=\sum_{k=0}^{\infty} \xi(z-k)\frac{z^k}{k!}+\Gamma(1-s)(-z)^{s-1},
\end{align}
for $|z|<2\pi$. Therefore, if parameter $s$ takes large negative values, the first sum in expression (\ref{eq:approx}) disappears as the Riemann zeta function $\xi(\cdot)$ approaches fast to $0$ for large negative values of $s$, and thus $\textrm{F}(z,s) \approx \Gamma(1-s)\left(-\log(z)\right)^{s-1}$. We have approximated the probability mass function given in expression (\ref{eq6}) for extreme cases (e.g., $s< -120$) as given below:
\begin{align}\label{eq:probs2}
\textrm{P}(X=x)\approx\exp\left(\log(z)\left(x+1\right)-s\log\left(x+1\right)-\log\left(\Gamma(1-s)\right)-\left(s-1\right)\log\left(-\log(z)\right)\right).
\end{align}

\subsection{Cumulative distribution function}

The cdf of the Good distribution is implemented in \texttt{good} through the function \texttt{pgood}. This function can be called as given below:

\begin{example}
pgood (q , z , s , lower.tail = TRUE )
\end{example}

The description of these arguments is summarized as follows:
\begin{itemize}
\item \texttt{q}: vector of non-negative integer quantiles.
\item \texttt{lower.tail}: Logical; if \texttt{TRUE} (default), probabilities are $\textrm{P}(X \leq x)$; otherwise, $\textrm{P}(X > x)$.
\end{itemize}

All remaining arguments $z$ and $s$ are defined the same as for the function \texttt{dgood}. Notice that as the function \texttt{pgood} internally calls the function \texttt{dgood}, the cdf is also approximated for cases where the parameter $s$ takes large negative values with the approximation described in expressions (\ref{eq:approx}) and (\ref{eq:probs2}). 

\subsection{Quantile function}

The quantile function of the Good distribution is implemented in the package \texttt{good} through the function \texttt{qgood}. This function can be called as given below:

\begin{example}
qgood (p , z , s , lower.tail = TRUE )
\end{example}

The description of these arguments is summarized as follows:
\begin{itemize}
\item \texttt{p}: vector of probabilities.
\end{itemize}

All remaining arguments are defined as specified for the functions \texttt{dgood} and \texttt{pgood}. The quantile is right-continuous: The function \texttt{qgood ( p , z , s ) } gives the smallest integer $x$ such that $\textrm{P}(X \leq x) \ge p$, where $X$ is a random variable following a Good distribution with parameters $z$ and $s$. 

\subsection{Random number generation}

The random number generation of the Good distribution is implemented in the package \texttt{good} through the function \texttt{rgood}. This function can be called as given below:

\begin{example}
rgood ( n , z , s , th )
\end{example}

The description of these arguments is summarized as follows:
\begin{itemize}
\item \texttt{n}: vector of number of observations to be generated. 
\item \texttt{th}: defines the lower ($q_1$) and upper ($q_2$) quantiles such that $\textrm{P}(X \leq q_1)=\textrm{th}$ and $\textrm{P}(X \leq q_2)=1-\textrm{th}$ respectively. Default $\textrm{th}=10^{-6}$.
\end{itemize}

All remaining arguments are defined as specified for the function \texttt{dgood}. Notice that we use here the cdf of a Good distribution to draw random samples from a population following such a distribution. In particular, we compute the cdf of a Good distribution with parameters $z$ and $s$ by means of the function \texttt{pgood} over the vector of quantiles with minimum and maximum $q_1$ and $q_2$ respectively. These quantiles are defined in so that $\textrm{P}(X \leq q_1)=\textrm{th}$ and $\textrm{P}(X \leq q_2)=1-\textrm{th}$, where $X \sim \textrm{Good}(z,s)$ and the argument $\textrm{th}$ is $10^{-6}$ by default. Finally, given a random value $u \sim \textrm{Uniform}(0,1)$, the corresponding cdf of the $X \sim \textrm{Good}(z,s)$ and the associated quantiles, the random value for $X$ will be the smallest integer $x$ such that $\textrm{P}(X \leq x)\geq u $. Of course, such an integer $x$ will lie within the interval $(q_1, q_2)$. 

\subsection{Maximum likelihood estimation and Good regression}

Assume the random variables $X_1, \cdots, X_n$ are independent copies of the random variable $X \sim \textrm{Good}(z,s)$ with pmf given in equation (\ref{eq9.5}), where parameters $z$ and $s$ are unknown and thus have to be estimated. Suppose $x_1, \cdots, x_n$ are realizations of  $X_1, \cdots, X_n$ (and thus $X$). The log-likelihood function of the Good distribution with parameters $z$ and $s$ is given by: 
\begin{equation} \label{lik1}
\textrm{l}\left(x_1,\cdots,x_n|z,s\right)=-n \log\left(\textrm{F}\left(\log(z),s\right)\right)+\log(z)\left(\sum_{i=1}^n (x_i+1)\right)-s\sum_{i=1}^n \log(x_i+1),
\end{equation}

which was previously given by \cite{DL97}. In fact, the authors showed that $(\bar{X},\bar{X}^{*})$, which are respectively the arithmetic and geometric means, is a joint sufficient statistic for the vector of parameters $(\alpha,\beta)$. In our parametrization above, parameters $\alpha$ and $\beta$ equate $\alpha=\log(z)$ and $\beta=-s$. The authors also pointed out that the joint sufficient statistic above is also a complete statistic as the Good distribution belongs to the exponential family, and studied the asymptotic properties of the maximum likelihood estimators (MLE) for $z$ and $s$. 

The package \texttt{good} includes the function \texttt{glm.good} for parameter estimation and Good regression. In particular, we assume the response variable $X_i \sim \textrm{Good}(z_i,s)$ for $i=1, \cdots, n$, where the parameter $s$ is fixed, and the parameter $z_i=h^{-1}( \boldsymbol{x_i^t} \beta)$ is related to the vector of covariates such that $\boldsymbol{x_i^t}$ is the transpose vector of covariates for the i$^{\textrm{th}}$ observation, $\beta=\left(\beta_0, \beta_1, \cdots, \beta_p\right)$ is the corresponding vector of coefficients to be estimated by maximum likelihood, and $h$ is the link function that can either be the identity, log, or logit. Since the Good distribution cannot be easily re-parametrized in terms of e.g., the parameter $\mu$ (expectation), it is thus not possible to link the parameter $\mu$ with the vector of predictors as usually done in e.g., the Poisson regression. Therefore, we have decided to include covariates in the regression through the parameter $z$. 

The log-likelihood function in expression (\ref{lik1}) is numerically maximized with respect to the parameters of interest. If we use both the log and identity link functions, we can obtain estimates for the coefficients $\beta$ (recall that $z_i=h^{-1}({\bf x}^t_i\beta)$) such that the estimated values for the parameter $z$ lie outside its parameter space, i.e., $(0,1)$. To overcome such a problem, we have to specify appropriate constraints on the $\beta$ coefficients ensuring the resulting estimated $z_i$ are within the corresponding parameter space (i.e., $z_i \in (0,1)$). In particular, we use the function \texttt{nlm} from the package \texttt{base} to maximize the log-likelihood function if the logit identity link is considered (as no constraints over the coefficients $\beta$ are needed), and the function \texttt{maxLik} from the package \texttt{maxLik} (\cite{maxLik}) if both log and identity link function are considered as \texttt{maxLik} allows specifying constraints on the coefficients $\beta$. 

In practice, the functions above (i.e., \texttt{nlm} and \texttt{maxLik}) that numerically maximizes the log-likelihood function require initial values for the parameters. By default, the function \texttt{glm.good} uses different starting values depending on the specified link function. In particular, values $\log(0.5)$ and $0.5$ are considered for the intercept ($\beta_0$) if respectively the log and identity link functions are considered, and $0$ is considered as starting values for the coefficients related to the covariates ($\beta_j$ for $j=1,\cdots, p$) for both link functions. Notice that if such initial values are used, we thus initialize functions \texttt{nlm} and \texttt{maxLik} assuming $z_i=0.5$ for $i=1, \cdots, n$, which is the less informative option as the parameter $z$ lies within $(0,1)$. However, if the logit link function is specified, we consider $\beta_0=\beta_1=\cdots=\beta_p=0$. In such a case, we again initialize functions \texttt{nlm} and \texttt{maxLik} assuming $z_i=\exp(0)/(1+\exp(0))=0.5$ for $i=1, \cdots, n$. For all three link functions, parameter $s$ is initialized at $-2$ as this value is roughly the boundary between under-dispersion (below $s \ll -2$) and over-dispersion (above $s \gg -2$).

The function \texttt{glm.good} can be called as given below: 
\begin{example}
glm.good ( formula , data , link = "log" , start = NULL )
\end{example}
The description of these arguments is summarized as follows:
\begin{itemize}
\item \texttt{formula}: Symbolic description of the model to be fitted. A typical predictor has the form \texttt{response $\sim$ terms} where \texttt{response} is the (integer) response vector and \texttt{terms} following a Good distribution with parameters $s$ and $z$, and \texttt{terms} is a series of covariates.
\item \texttt{data}: An optional data frame containing the variables in the model.
\item \texttt{link}: Character specification of link function: \texttt{"log"}, \texttt{"logit"} or \texttt{"identity"}. By default \texttt{link = "log"}.
\item \texttt{start}: A vector with the starting values for the model parameters. Used for numerically maximize the likelihood function for parameters estimation. Its default value is \texttt{NULL}.
\end{itemize}

The function returns an object of class \texttt{glm.good} that is a list with following components:
\begin{itemize}
\item \texttt{coefs}: The vector of estimated coefficients.
\item \texttt{loglik}: Log-likelihood of the fitted model.
\item \texttt{vcov}: Variance-covariance matrix of all model parameters (derived from the Hessian returned by \texttt{nlm()} output). 
\item \texttt{hess}: Hessian matrix, returned by the \texttt{nlm()} or \texttt{maxLik()} output.
\item \texttt{fitted.values}: The fitted mean values. These are obtained by transforming the linear predictors with the inverse of the link function. 
\end{itemize}

For an object of type \texttt{glm.good}, we can use the functions \texttt{summary} and \texttt{predict}. In particular, the function \texttt{summary} returns relevant information concerning the estimated Good regression, such as a naive description of the empirical distribution of the residuals (minimum, maximum, median and first and third quartiles), the table of estimated coefficients with standard errors and the asymptotic Wald significance test for each of the parameters $\beta_j$ for $j=0, \cdots, p$, and if no covariates are considered, the point estimate and standard error (with the univariate Delta method) of parameter $z$. The function \texttt{summary} over an object \texttt{glm.good} also returns some goodness-of-fit measures and results for some significance tests. In particular, as goodness-of-fit measures, the function returns both the Akaike Information Criterion (AIC) and Bayesian Information Criterion (BIC) values. In addition, if no covariates are considered in the regression, we obtain the results of the following Likelihood Ratio Test (LRT) : (i) $\textrm{H}_0: \textrm{Logarithmic distribution} \, (s=1)$ vs. $\textrm{H}_1: \textrm{Good distribution} \, (s\ne 0)$, and (ii) $\textrm{H}_0: \textrm{Geometric distribution} \, (s=0)$ vs. $\textrm{H}_1: \textrm{Good distribution} \, (s\ne 0)$. However, if covariates are considered in the regression, results for the following LRT are given: $\textrm{H}_0: \textrm{Model without covariates} \, (\beta_1=\cdots=\beta_p=0)$ vs. $\textrm{H}_1: \textrm{Model with covariates} \, (\beta_j\ne 0, \, \exists j \, \textrm{for} \, j=1, \cdots p)$. 

\subsection{Predictions}\label{predictions}

Predictions can also be computed for an object of class \texttt{glm.good} as the function \texttt{predict} has been adapted for objects of such a class. In particular, we can predict the mean values $\mu_i$ for $i=1, \cdots, n$ and their standard errors for a Good distribution with estimated parameters $\widehat{z}_i$ for $i=1, \cdots, n$ and $\widehat{s}$. Considering the expression (\ref{eq9}), a point estimate for $\mu_i$ can be easily computed such that:
\begin{align}\label{eq:pointest}
\widehat {\mu}_i = \frac{\textrm{F}(\widehat{z}_i(\widehat{\beta} ),\widehat{s}-1)}{\textrm{F}(\widehat{z}_i(\widehat{\beta}),\widehat{s})}-1, \quad \textrm{for} \quad i=1, \cdots, n, 
\end{align}

where $\widehat{z}_i(\widehat \beta)=h^{-1}({\bf x}_i^{\textrm{t}}\widehat {\beta})$ and ${\bf x}_i^{\textrm{t}}\widehat{\beta}=\left(\widehat{\beta}_0+\widehat{\beta}_1 x_{i1}+\cdots+\widehat{\beta}_p x_{ip}\right)$. Notice that if no covariates are given, then $\beta_1 =\beta_2 =\cdots =\beta_p=0$ and therefore $\widehat{z}_i(\widehat{\beta})=h^{-1}(\widehat{\beta}_0)$ which is constant for $i=1, \cdots, n$. Recall that $h^{-1}(\cdot)$ is the inverse of the link function (identity, log or logit). 

The standard errors of the estimates in (\ref{eq:pointest}) can be computed with the multivariate Delta method. Formally, consider a vector of parameters $\Theta=(\theta_1, \cdots, \theta_p)$, and a vector $T=\left(T_1(x), \cdots, T_k(x)\right)$ for $k \leq p $ of estimators  for $\Theta$. Consider also the transformation $\eta=g(\Theta)$ such that $\partial g (\theta_j)/\partial \theta_j \ne 0$ for $j=1, \cdots, p$. Therefore, the approximate variance of $\eta$ is given by: $\textrm{Var}(\eta) = \nabla g(\Theta)^{\textrm{T}} \Sigma  \nabla g(\Theta)$, where $ \nabla g(\Theta)^{\textrm{T}}=\left(\partial g(\Theta)/\partial \theta_1, \cdots, \partial g(\Theta)/\partial \theta_p\right)$ and $\Sigma$ is the variance-covariance matrix of estimators of $\Theta$.

In the case here, we have a vector of parameters given by $\Theta=(s,{\beta})$ where ${\beta}=(\beta_0, \beta_1, \cdots, \beta_p)$ such that $z_i(\widehat{\beta})=h^{-1}({\bf x}_i^{\textrm{t}}{\beta})$ for $i=1, \cdots, n$. Our vector of estimators is given by the MLE of $\Theta$ such that $T=(\widehat{s},\widehat{\beta}_0, \widehat{\beta}_1, \cdots, \widehat{\beta}_p)$, and $g_i(\Theta)=g_i\left(\left(s, \beta\right)\right)=\textrm{F}(z_i(\beta),s-1)/\textrm{F}(z_i(\beta),s)-1$, given in expression (\ref{eq:pointest}).

If we take the first derivatives of $g_i\left(\left(s, \beta\right)\right)$ with respect to parameters $s$ and $\beta$, vector $\nabla g_i\left(\left(s, \beta\right)\right)^{\textrm{T}}$ is given by:
\begin{align}\label{eq:eq4}
\nabla g_i\left(\left(s,\beta\right)\right)^{\textrm{T}}&=\left(\frac{\partial g_i\left(\left(s,\beta\right)\right)}{\partial s}, \frac{\partial g_i\left(\left(s,\beta\right)\right)}{\partial \beta_0},\cdots, \frac{\partial g_i\left(\left(s,\beta\right)\right)}{\partial \beta_p}\right)^{\textrm{T}},
\end{align}

where $\partial g_i\left(\left(s,\beta\right)\right)/\partial s$ depends on $\partial \textrm{F}(z_i(\beta),s)/\partial s$ and $\partial \textrm{F}(z_i(\beta),s-1)/\partial s$, and $\partial g_i\left(\left(s,\beta\right)\right)/\partial \beta_j$ for $j=0, \cdots, p$ depends on $\partial \textrm{F}(z_i(\beta),s)/\partial \beta_j$, and $\partial \textrm{F}(z_i(\beta),s-1)/\partial \beta_j$ for $j=0,\cdots,p$. 

Therefore, one can easily see that $\partial \textrm{F}(z_i(\beta),s)/\partial s=-\sum_{n=1}^{\infty} (z_i^n\log(n))/n^{s}$ and $\partial \textrm{F}(z_i(\beta),s-1)/\partial s=-\sum_{n=1}^{\infty} (z_i^n\log(n))/n^{s-1}$. 

However, the computations of $\partial F(z_i(\beta),s)/\partial \beta_j$ and $\partial F(z_i(\beta),s-1)/\partial \beta_j$ for $j=0,\cdots,p$ slightly vary depending on the specified link function. Therefore, if we consider the identity link function, $\partial \textrm{F}(z_i(\beta),s)/\partial \beta_j=(x_{ij}\textrm{F}(z_i(\beta),s-1))/z_i$ and $\partial \textrm{F}(z_i(\beta),s-1)/\partial \beta_j=(x_{ij}\textrm{F}(z_i(\beta),s-2))/z_i $. If we used instead a log link function, $\partial \textrm{F}(z_i(\beta),s)/\partial \beta_j=(x_{ij}\sum_{n=1}^{\infty} z_i(\beta)^n)/n^{s-1}$ and $\partial \textrm{F}(z_i(\beta),s-1)/\partial \beta_j=(x_{ij}\sum_{n=1}^{\infty} z_i(\beta)^n)/n^{s-2}$ Finally, if the logit link function is considered, $\partial F(z_i(\beta),s)/\partial \beta_j=x_{ij} \sum_{n=1}^{\infty}(z_i(\beta)^n)/(n^{s-1}(1+\exp(z_i^*(\beta))))$ and $\partial F(z_i(\beta),s-1)/\partial \beta_j=x_{ij} \sum_{n=1}^{\infty}(z_i(\beta)^n)/(n^{s-2}(1+\exp(z_i^*(\beta))))$, where $z_i^*(\beta)=\textrm{logit}(z_i(\beta))$; all three cases for $j=0,\cdots, p$. Notice here that if we use log or logit link functions $z_i(\beta)=\exp(\beta_0+\beta_1x_{i1}+\cdots+\beta_p x_{ip})$ or $z_i(\beta)=1/(1+\exp(-\beta_0-\beta_1x_{i1}-\cdots-\beta_p x_{ip}))$, respectively. However, if the link function is the identity, $z_i(\beta)=\beta_0+\beta_1 x_{i1}+\cdots+\beta_p x_{ip}$. 

Replacing $\nabla g_i\left(\right(s,\beta\left)\right)^{\textrm{T}}$ by $\nabla g_i\left(\right(\widehat s,\widehat \beta\left)\right)^{\textrm{T}}$, we can compute the variance of $\widehat{\mu}_i$ in expression (\ref{eq:pointest}) such that: 
\begin{align}
\widehat{\textrm{Var}}(\widehat{\mu}_i)= \nabla g_i\left(\right(\widehat s,\widehat \beta\left)\right)^{\textrm{T}} \widehat{\Sigma}  \nabla g_i\left(\right(\widehat s,\widehat \beta\left)\right),
\end{align}

where the estimated variance-covariance matrix $ \widehat{\Sigma} $ can be numerically obtained with both the functions \texttt{maxLik} and \texttt{nlm} used in \texttt{glm.good} for maximum likelihood estimation.  
The package \texttt{good} returns point estimates and standard errors with the function \texttt{predict}.

\section{Examples} \label{exp}

Several examples of application for the \texttt{good} package are discussed below.

\subsection{McNeil (1975)}  \label{ex1}

This example by \cite{Mc77} analyses the yearly number of "great" inventions and scientific discoveries from 1960 to 1959. The data are available in the \texttt{R} package \texttt{datasets}. The sample mean and variance are respectively $3.10$ and $5.08$, with a sample dispersion index of $d = 1.64$, showing therefore over-dispersion with respect to the Poisson distribution. The Negative Binomial models are very often used for modelling over-dispersed count data. In the current example, we fit the number of great inventions and discoveries with the Negative Binomial (NB) and Good distributions, both allowing for over-dispersion. 

\begin{table}[h!]
\centering
    \begin{tabular}{ l ccccc cc} 
     \toprule
     Inventions &    0 &     1 &     2 &     3 &    4 &    5 &    6  \\ \midrule
     Observed   &    9 &    12 &    26 &    20 &   12 &    7 &    6  \\
     NB         & 8.59 & 16.98 & 19.86 & 17.88 & 13.70 & 9.39 & 5.93 \\
     Good       & 7.73 & 17.82 & 20.59 & 17.92 & 13.36 & 9.03 & 5.71 \\ \midrule
     
     Inventions &   7  &   8  &    9 &   10 &   11 &   12 &  13+  \\ \midrule
     Observed   &   4  &   1  &    1 &    1 &    0 &    1 &    0  \\
     NB         & 3.51 & 1.98 & 1.07 & 0.56 & 0.29 & 0.14 & 0.13  \\
     Good       & 3.43 & 1.99 & 1.12 & 0.61 & 0.33 & 0.17 & 0.19  \\
     \bottomrule
    \end{tabular}
\caption{Observed frequencies and expected frequencies under NB and Good distributions for the number of "great" inventions by year.}
\label{tab1}
\end{table}

Table \ref{tab1} shows the observed and expected frequencies under the Negative Binomial and Good distributions. We fit the Negative Binomial regression (without covariates) with the function  \texttt{glm.nb} from the package \texttt{MASS} (\cite{MASS}). We have obtained an AIC and BIC of $425.45$ and $430.66$ for the Good distribution respectively, and $425.59$ and $430.80$ for the NB distribution respectively.   

With the function \texttt{glm.good}, we have estimated parameters of a Good distribution. A summary of the most relevant information can be obtained using the function \texttt{summary()} whose output is shown below:

\begin{example}
    > library ( "good" )
    > data ( "discoveries" )
    > fit.discoveries <- glm.good ( discoveries ~ 1 , link = "log" )
    > summary ( fit.discoveries )
    Call:
    glm.good ( formula = discoveries ~ 1 , link = "log" )
        
    Deviance Residuals:
        Min         1Q      Median        3Q        Max 
    -3.1002074 -1.1002074 -0.1002074  0.8997926  8.8997926 
    --
    Coefficients:
                  Estimate  Std. Error  z value   p-value
    s           -2.4022100  0.4940341 -4.862438 1.15949e-06
    (Intercept) -0.8295909  0.1275701 -6.503022 7.87225e-11
    --
    Transformed intercept-only parameter
       Estimate Std. Error
    z 0.4362277  0.0556496
    --
    Likelihood ratio test:
    Model 1: logarithmic (s=1)
    Model 2: good 
      #Df  LogLik Df     LRT    p.value
    1   1 -252.45                      
    2   2 -210.73  1 83.4392 < 2.22e-16
        
    Model 1: geometric (s=0)
    Model 2: good 
      #Df  LogLik Df     LRT    p.value
    1   1 -227.77                      
    2   2 -210.73  1 34.0861 5.2725e-09
    --
    LogLik: -210.73    AIC: 425.45    BIC: 430.66 
\end{example}

As the log link function is specified, the point estimate of the parameter $z$ is thus $\hat{z}=\exp(-0.8295)=0.4362$ with an associated standard error of $0.0556$ (Transformed intercept-only parameter), which is easily derived from the Delta method given that $\widehat{\sigma}^2_{\widehat{z}} \approx \widehat{\sigma}^2_{\widehat{\eta}}\left(\exp(\widehat{\eta})\right)^2$ and $\widehat{\eta}=\log(\widehat{z})$. The point estimate of the parameter $s$ is $\hat{s}=-2.4022$ with an associated standard error of $0.4867$. We can easily see that the expected mean and variance of the $\textrm{Good}(0.4362,-2.4022)$ are respectively $3.10$ and $4.94$, which are not far away from the sample mean ($3.01$) and variance ($5.08$). Finally, results of the likelihood ratio significance tests show that the Good distribution models the data more appropriately compared to either the Logarithmic ($s=1$) or Geometric ($s=0$) distributions.  

It can be seen that both models work considerably well. However, the Good distribution seems to work slightly better; even the differences in the AIC and BIC values can be considered irrelevant. Despite that, this example illustrates that the Good distribution can work similarly (or better) than the typical models (e.g., Negative Binomial distribution) for addressing over-dispersion in real-world data configurations.

\subsection{Kendall (1961)}  \label{ex2}
 
The data set here corresponds to the number of outbreaks of strikes in 4-week periods in a coal mining industry in the United Kingdom during 1948-1959. This data set has been previously analysed by other authors such as \cite{K61}, \cite{C98} and \cite{W13a}. The sample mean and variance are respectively $0.994$ and $0.742$, with a dispersion index of $d = 0.75$, showing under-dispersion. As this data set has been analysed by other authors who proposed alternative models allowing under-dispersion, we will compare some of these models to the Good distributions. In particular, we will use the Poisson distribution, the GP distribution, the power-law distribution (PL$_{\nu}$) (\cite{C98}), the COM-Poisson distribution and the Good distribution. Notice that \cite{W13a} already used the Good distribution for this particular example for modelling count time series showing under-dispersion.

The three-parameter power law-distribution, as shown by \cite{C98}, addresses under-dispersion for values of the parameter $\nu$ greater than $0$. We particularly follow here the parametrization proposed by \cite{W13a} and consider the same two scenarios of under-dispersion with $\nu=1$ and $\nu=2$. Notice that the Good distribution has one parameter less than the power-law distribution but still addresses both under-dispersion and over-dispersion in a rather flexible manner. Recall also that the two-parameter GP and COM-Poisson distributions address under-dispersion for values of the parameter $\lambda $ lower than $0$ and $\nu$ greater than $1$ respectively. 

The sample mean and sample variance are $0.994$ and $0.742$, respectively, with a dispersion index of $d = 0.747$, showing thus under-dispersion. Table \ref{tab2} shows both the observed and predicted frequencies under each model considered above. Table \ref{tab3} shows the maximum likelihood point estimates (and associated standard errors) for each model's parameters and the AIC and BIC values as goodness-of-fit measures. Although the power-law and Good distributions have similar results concerning AIC and BIC, we see that the Good distribution is the one with the smallest AIC and BIC. In particular, we obtain estimates $\hat{z}=0.057 \, (0.022)$ (with the log link function) and $\hat{s}=-4.776 \, (0.721)$, which lead to a Good distribution with expectation $0.994$, variance $0.730$ and dispersion index $0.735$, quite similar to the empirical ones given above. As in the previous example, we have used the function \texttt{glm.good} (and its link to the function \texttt{summary}) to obtain the results displayed in Table \ref{tab3}. Notice that, unlike the Good distribution, the power-law distributions $\textrm{PL}_1$ and $\textrm{PL}_2$ are three-parameter. Naturally, we expect such distributions to model better the data than a two-parameter distribution such as the Good distribution. Therefore, results given in Table \ref{tab3} support, even more, the conclusion that the Good distribution is the best option (among the ones considered in the current example) to model the number of outbreaks of strikes in 4-week periods in a coal mining industry in the United Kingdom during 1948-1959.
 
 \begin{table}[h]
 \centering
     \begin{tabular}{ cccccccc } \toprule
      Outbr. & Obs & Poisson & GP & PL$_1$ & PL$_2$ & COM-Poisson & Good \\ \midrule 
      0       & 46 & 57.76 & 50.01 & 46.27 & 46.58 & 47.49 & 46.64 \\ 
      1       & 76 & 57.39 & 65.77 & 73.16 & 72.29 &  70.43 & 72.86 \\ 
      2       & 24 & 28.51 & 32.23 & 29.09 & 29.77 & 30.66 & 28.79 \\ 
      3       & 9  & 9.44  & 7.23  & 6.39  & 6.37  & 6.51 & 6.48  \\ 
      4+      & 1  & 2.90  & 0.76  & 1.09  & 0.99  & 0.91 & 1.24  \\ 
    \bottomrule
    \end{tabular}
 \caption{Observed frequencies and expected frequencies under the Poisson, GP,  PL$_1$, PL$_2$, COM-Poisson and Good distributions for the number of strikes in a coal mining industry in the United Kingdom from 1948 to 1959.}
 \label{tab2}
 \end{table}
 \begin{table}[h]
 \centering
     \begin{tabular}{ lccccccr } \toprule
                    &                     & Par. 1         & Par. 2          &   Par. 3 &AIC    & BIC    \\ \midrule
      Poisson       & ($\lambda$)         & 0.994 (0.080)  &                & & 385.87 & 388.92 \\ 
      GP            & ($\lambda, \theta$) &-0.145 (0.055)  & 1.138 (0.106) &  & 381.65 & 387.74 \\ 
      PL$_1$        & ($\lambda$, $a$, $\nu$)       & 0.466 (0.085)  & 0.419 (0.187)  & $1^{*}$ & 379.20 & 385.30 \\ 
      PL$_2$        & ($\lambda$, $a$, $\nu$)       & 0.356 (0.080)  & 0.919 (0.080)  & $2^{*}$ & 379.52 & 385.62 \\ 
      COM-Poisson   &       ($\lambda$, $\nu$)      &  1.483 (0.252) &  1.769 (0.295)  & &  380.02 &  386.12\\  
      Good          & ($z$, $s$)            & -2.865 (0.370) & -4.776 (0.721) & & 379.14 & 385.24 \\ \bottomrule
    \end{tabular}
 \caption{Maximum Likelihood Estimates (MLE) and associated standard errors for the parameters of the Poisson, GP, PL$_1$, PL$_2$, COM-Poisson and Good distributions for the number of strikes in a coal mining industry in the United Kingdom from 1948 to 1959.\\ $^{*}$Indicates that the corresponding parameter has considered known and particularly with values $1$ and $2$.}
 \label{tab3}
 \end{table}

\subsection{Folio et al. (2019)}  \label{exPolar}

Data here are the litter size of polar bears from $1992$ to $2017$ at Svalbard, Norway, from late March to the beginning of May analysed by \cite{D19}. The sample mean and variance are $1.705$ and $0.278$, respectively, with a dispersion index of $d = 0.163$ and thus showing a severe under-dispersion. As these data are zero-truncated, an appropriate parametric model could be the zero-truncated Poisson (ZTP) distribution. We also consider both the GP and COM-Poisson distributions as some of the most typical alternatives for addressing under-dispersion and, finally, the Good distribution. The observed and expected frequencies under the ZTP, GP, COM-Poisson and Good distributions are given in Table \ref{tabpolar}. We provide point estimates and standard errors in Table \ref{tabpolar2} and also both the AIC and BIC values as model selection measures. We can see in Table \ref{tabpolar2} that the Good distribution works significantly better than the ZTP and GP, obtaining the following estimates for the parameters $z$ and $s$: $\hat{z}=\exp(-11.671)=8.538 \times 10^{-6}\, (2.271 \times 10^{-6})$ and $\hat{s}=-30.413 \, (0.613)$, with an expected mean and variance of respectively $1.706$ and $0.277$ and dispersion index $ 0.163$, all really close to the empirical ones. Notice that the Good distribution shows better AIC and BIC values compared to the COM-Poisson distribution, although the differences are really tiny. 
\begin{table}[h!]
\centering
    \begin{tabular}{ l ccccc ccccc ccccc ccccc ccccc cccc } 
     \toprule
     Litter Size    &   0  &     1 &     2  &     3 &     4+ \\ \midrule
     Observed &   0  &    76 &   147  &     8 &     0  \\ 
     ZTP      &  0* & 120.68 &  71.39 & 28.16 & 10.77  \\ 
     GP       & 6.29 & 68.96 & 144.55 &  7.51 & 0*     \\ 
     COM-Poisson & 0.08 & 75.68 & 147.43 & 7.77  &  0.03    \\ 
     Good     & 0.01 & 76.00 & 147.02 &  7.91 &  0.06  \\ 
    \bottomrule
    \end{tabular}
\caption{Observed frequencies and expected frequencies under ZTP, GP, COM-Poisson and Good distributions for the number of polar bears in a litter. The * symbol means that the corresponding value is truncated by definition of the distribution.}
\label{tabpolar}
\end{table}
\begin{table}[h!]
 \centering
     \begin{tabular}{ lc cccc } 
     \toprule
                    &                     & Par. 1          & Par. 2          & AIC    & BIC    \\ \midrule
      ZTP           & ($\lambda$)         &  1.183 (0.086)  &                 & 479.57 & 483.01 \\ 
      GP            & ($\lambda, \theta$) & -1.112 (0.074)  & 3.603 (0.198)   & 378.38 & 381.82 \\ 
      COM-Poisson          & ($\lambda$, $\nu$)            &  931.83 (697.50) & 8.90 (0.975) & 359.90 & 366.79 \\ 
      Good          & ($z, s$)            & -11.671 (0.266) &-30.413 (0.613)  & 359.80 & 366.69 \\ \bottomrule
    \end{tabular}
 \caption{Maximum Likelihood Estimates (MLE) and associated standard errors for the ZTP, GP, COM-Poisson and Good distributions for the number of polar bears in a litter.}
 \label{tabpolar2}
 \end{table}
 
\subsection{Fraile et al. (2020)}  \label{expPiglets}

Data here are based on a previous work by \cite{F19} who developed a new method to detect re-circulation of Porcine Reproductive and Respiratory Syndrome (PRRS) in sow production farms with a conditional Poisson model of the number of lost piglets on the number of born alive piglets. We focus in the following on the number of born alive piglets (NBA) with a sample mean and variance of $13.261$ and $11.698$ respectively, and a dispersion index of $0.882$. We particularly aim at regressing NBA onto the number of sows (\texttt{parity}) through the function \texttt{glm.good} as shown below:

\begin{example}
    > fit.piglets <- glm.good(nba ~ parity, data = piglets, link = "logit")
\end{example}

We here used the logit link function as compared to the other examples where we used the log link function between parameter $z$ and predictors. We show below the output from \texttt{summary()} and \texttt{predict()} applied to an object \texttt{glm.good}:
\begin{example}
    > summary(fit.piglets)
    Call: glm.good ( formula = nba ~ parity , link = "logit" )
    --
    Deviance Residuals:
        Min           1Q        Median        3Q         Max
    -13.2249051  -2.0104247   0.5559896   2.1300469   9.6295454 
    --
    Coefficients:
                    Estimate  Std. Error     z value     p-value
    s           -12.474133071 0.390905463 -31.910869 1.886972e-223
    (Intercept)  -0.425058084 0.047212040  -9.003171  2.192898e-19
    parity       -0.007859734 0.003721204  -2.112148  3.467373e-02
    --
    Likelihood Ratio Test:
    Model 1: nba ~ 1
    Model 2: nba ~ parity 
    #Df   LogLik Df    LRT  p.value
    1   1 -6381.34                   
    2   2 -6379.11  1 4.4621 0.034655
    --
    LogLik: -6379.11    AIC: 12764.21    BIC: 12781.46 
    
    > pred.discoveries <- predict ( discoveries , se.fit = TRUE )
    > names(discoveries)
    [1] "fit"    "se.fit"
    > head ( pred.discoveries [[ 1 ]] , 3 )
    [1] 12.66294 12.59490 12.52735 
    > head ( pred.discoveries [[ 2 ]] , 3 )
    [1] 0.2849865 0.3140257 0.3428341 
\end{example}

\begin{figure}[h!]
    \centering \includegraphics[width=14cm]{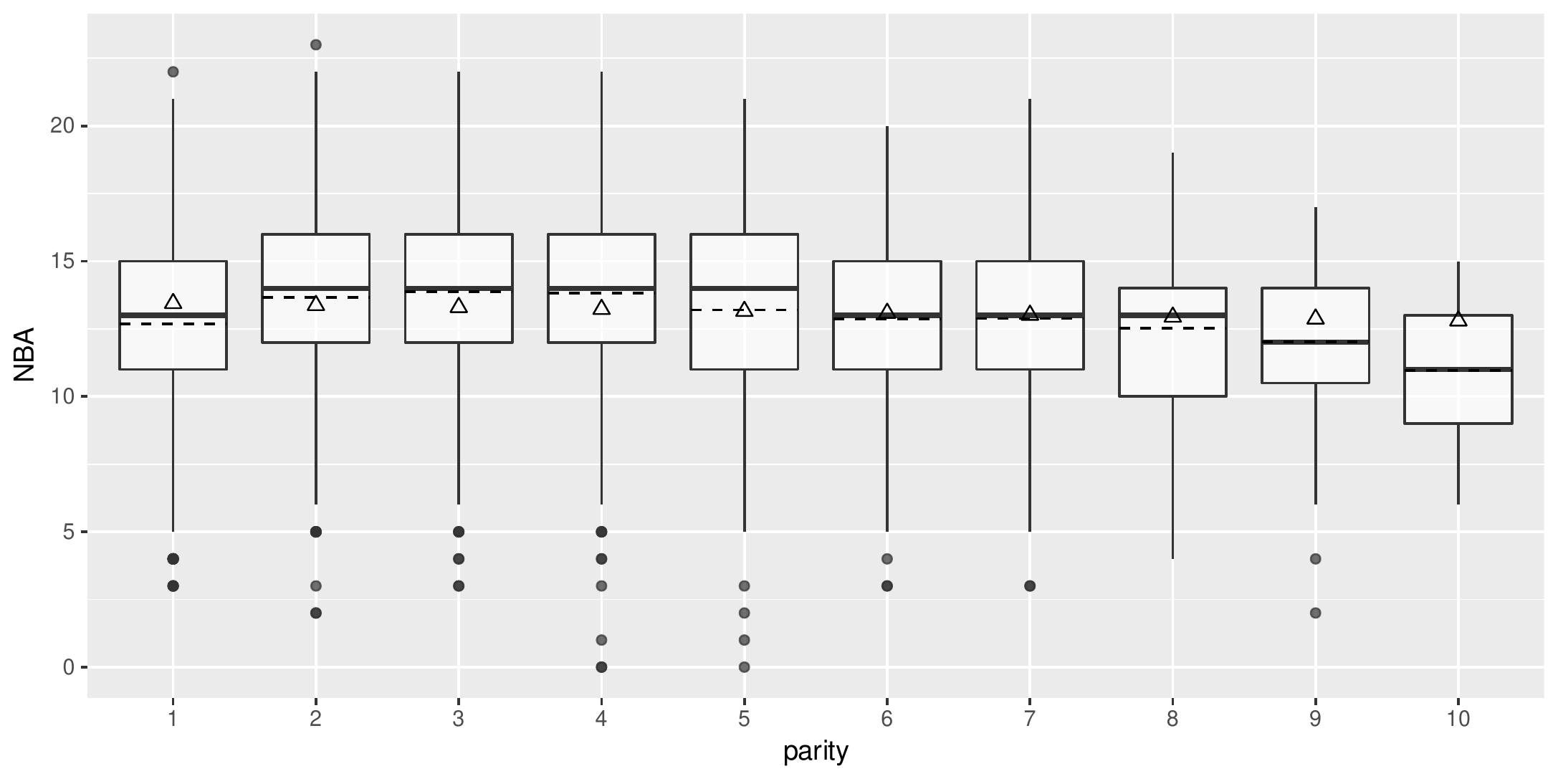}
    \vspace{-0mm}
    \caption{Empirical distributions of the number of alive piglets by parity. The triangle symbol indicates the predicted mean of the number of alive piglets for each sub-group (by parity) under the Good distribution with the estimated parameters given above. The dashed line indicates the observed mean of the number of alive piglets for each sub-group (by parity).}%
    \label{fig3}%
\end{figure}

The summary shows that parity is associated with the number of born alive piglets, and particularly we estimate $\hat{z}_i=1/\left(1+\exp\left(0.4250+0.0078\textrm{parity}_i\right)\right)$, and $\hat{s}=-12.47 \, (0.391)$, with corresponding standard errors given in the summary above. We can also use the Delta method here to obtain the standard error of $\hat{z}_i$ for a given value of the predictor parity. For example, if parity values are $1$ and $5$, we obtain $\widehat{z}_1=0.3935$ and $\widehat{z}_5=0.3860$, and their associated standard errors can be computed with the Delta method as shown in the previous examples. Figure \ref{fig3} shows the empirical distributions of the variable NBA depending on the value of the predictor parity, showing that predicted (triangle symbol) and observed (dashed line) means of NBA by parity are close in all parity cases except for values of parity over $10$, which are rather extreme. The summary above also returns results of the likelihood ratio test showing that we have evidence enough to reject the null hypothesis of $\textrm{H}_0: \beta_{\textrm{parity}}=0$, and therefore accept the alternative $\textrm{H}_1: \beta_{\textrm{parity}}\ne 0$, and conclude that here the model with the covariate \texttt{parity} is more likely for the data than the model without such a covariate. 

The function \texttt{predict} gives the mean predicted value for each parity value, and also the standard errors of such mean predicted values. These standard errors are computed with the Delta method described above in Section \ref{predictions}.

\section{Conclusions} \label{con}

The Good distribution has shown to be a rather flexible candidate for modelling under-dispersed count data, a phenomenon that researchers of particular fields might face very often. For example, the Good distribution seems appropriate for the error process of INteger AutoRegressive (INAR) models when the count time series shows under-dispersion (Wei\ss \,, 2013). Such a distribution has also shown good results for modelling applications in ecology (\cite{K92}) and sequential analysis (\cite{WP13}), among others. In the current paper, we have also demonstrated that the Good distribution gives sometimes better results in terms of fitting compared to the typically considered distributions for modelling under-dispersion such as the NB, GP, and COM-Poisson. 

We have presented in the following the \texttt{R} package \texttt{good}, which includes functions to compute main characteristics of the Good distribution (probabilities, cumulative probabilities and quantiles), and a function with a random generator for a Good distributed population. The package also incorporates a function for Good regression that allows including covariates of any kind -as it works in a very similar way to that of the base \texttt{glm} function - and includes some datasets of the examples presented in the following. 

\section{Acknowledgments}

This work was co-funded by Instituto de Salud Carlos III (COV20/00115), and RTI2018-096072-B-I00. A. Fernández-Fontelo acknowledges financial support from the German Research Foundation (D.F.G.). A. Arratia acknowledges support by grant TIN2017-89244-R from MINECO (Ministerio de Economía, Industria y Competitividad) and the recognition 2017SGR-856 (MACDA) from AGAUR (Generalitat de Catalunya).

\bibliographystyle{unsrt}  
\bibliography{references}  

\address{Jordi Tur\\
  Centre de Recerca Matemàtica\\
  Edifici C, Universitat Autònoma de Barcelona, Cerdanyola del Vallès, Bellaterra, Barcelona 08193\\
  Spain\\}
\email{jjtur@crm.cat}

\address{David Moriña\\
  Department of Econometrics, Statistics and Applied Economics, Riskcenter-IREA, Universitat de Barcelona\\
  Barcelona 08003\\
  Spain\\}
\email{dmorina@ub.edu}

\address{Pedro Puig\\
  Barcelona Graduate School of Mathematics (BGSMath)\\
  Departament de Matemàtiques, Universitat Autònoma de Barcelona\\
  Bellaterra, Barcelona 08193\\
  Spain\\}
\email{ppuig@mat.uab.cat}

\address{Alejandra Cabaña\\
  Barcelona Graduate School of Mathematics (BGSMath)\\
  Departament de Matemàtiques, Universitat Autònoma de Barcelona\\
  Bellaterra, Barcelona 08193\\
  Spain\\}
\email{acabana@mat.uab.cat}

\address{Argimiro Arratia\\
  Department of Computer Science, Universitat Politècnica de Catalunya\\
  Barcelona, 08034\\
  Spain\\}
\email{argimiro@cs.upc.edu}

\address{Amanda Fernández-Fontelo\\
  Chair of Statistics, School of Business and Economics, Humboldt-Universität zu Berlin\\
  Unter den Linden 6, Berlin 10099\\
  Germany\\}
\email{amanda.fernandez-fontelo@hu-berlin.de}

\end{document}